\documentclass[apl, pre, twocolumn, amsmath,amssymb]{revtex4}
\usepackage{graphicx}
\usepackage{subfigure}
\usepackage{amsfonts}
\usepackage{bm}
\usepackage{dcolumn}
\usepackage{color}

\begin{document}

\title{Hydrodynamic synchronisation  
 of  non-linear oscillators at low Reynolds number.}
\date{\today}

\author{M. Leoni$^1$ and T. B. Liverpool$^{1,2}$ }

\affiliation{$^1$Department of Mathematics, University of Bristol, Clifton, Bristol BS8 1TW, U.K. \\
$^2$ The Kavli Institute for Theoretical Physics, University of California, Santa Barbara, California 93106, USA}

\begin{abstract}
We introduce a generic  model of weakly non-linear self-sustained oscillator as a simplified tool to study synchronisation in a fluid at low Reynolds number.   
By  averaging over the fast degrees of freedom, we examine  the effect of hydrodynamic interactions on the slow dynamics of  two oscillators and show that they can lead to synchronisation.  Furthermore, we find that synchronisation is strongly enhanced  when  the oscillators are non-isochronous, which on the limit cycle means the oscillations have an amplitude-dependent frequency. Non-isochronity is determined by a nonlinear coupling $\alpha$ being non-zero.
 {We  find that  its ($\alpha$) sign determines if they synchronise in- or anti-phase.  }
  We then study  an infinite array of oscillators in the long wavelength  %and low frequency
 limit, in  presence of noise. 
For $\alpha > 0$,  hydrodynamic interactions can lead to a homogeneous synchronised state. 
Numerical simulations for a finite number of oscillators
confirm this
 %analytic findings 
 and, when  $\alpha <0$,
 show
  the propagation of  waves, reminiscent of metachronal coordination.
\end{abstract}
\maketitle

Collections of cilia and flagella  are  % another 
examples of systems that display synchronisation~\cite{Pikovsky}.
They  are microscopic active filaments attached to the  membrane of pro- and  eukaryote cells~\cite{Bray} whose synchronisation is thought to aid the efficiency of transport at the cellular scale. Typically arrays of cilia  generate fluid flows along tissues but can also be used, like  flagella, for the self-propulsion of swimming cells.
Due to their tiny size, the Reynolds number associated with these flows is negligible.
The coordinated beating of cilia is also thought to have important developmental implications, such as
the left-right symmetry breaking in the arrangement of the internal organs  in  the  early embryo~\cite{Nonaka}.  
A {precise} understanding of the role hydrodynamics plays in their synchronised motion,  is still missing.
% (detailed)
%(e.g. the relative importance of chemical and mechanical couplings)

Both cilium and flagellum are made of complex  subunits,  microtubules 
driven by molecular motors, and their modelling can be done at  many  levels.
As synchronisation takes place on length-scales larger than the individual filaments, to a first approximation the fine details of their internal structure can be ignored. This  coarse-grained approach has led to model  studies  of self-sustained oscillators ~\cite{CLJ03},  rotating beads~\cite{VJ2006, NEL2008,UG10}; beating filaments~\cite{GJ07}, 
as well as rigid rotating helices,~\cite{KP04,RS2005}. 
More recent work has focused on the conditions for hydrodynamic synchronisations for two oscillators~\cite{UG2011} and the phase dynamics of oscillators with long range interactions~\cite{Uchida}. Related experiments investigating the dynamics of micro-systems  have been performed in vivo on algae,~\cite{Polin, Gold}
and on simple model systems~\cite{kotar}, and even a macroscopic scale model of rotating paddles~\cite{Qian}. 
All these studies suggest that simple forms of active forces, e.g. as prescribed functions of time, are not enough to guarantee synchronisation. 
Rather, a complex,  non-linear relation between forces and velocities is necessary. 
Important questions therefore are what aspects of hydrodynamic interactions aid synchronisation and what features of oscillators make them good hydrodynamic synchronizers.
%Their findings are consistent with a purely mechanical hydrodynamic source of  synchronisation.

The dynamics of a system close to an oscillatory instability can be conveniently described by weakly non-linear oscillators whose averaged equations are universal~\cite{Pikovsky}. This implies that the long time behaviour of many systems with {\em simple} spontaneous oscillations can be captured by a generic model with a few parameters. Using this insight, in this paper we introduce a minimal model of an oscillator at low Reynolds number.
To simplify our presentation, we study our model in one-dimension.
At a coarse grained level, this degree of  freedom can be interpreted as the centre of a filament  beating in a plane~\cite{Camalet99}.

The slow dynamics of the oscillator is naturally characterised using  of two variables: the amplitude and the phase.
Under arbitrary initial conditions, the trajectories of an isolated oscillator on long timescales converge to a closed curve, the limit cycle~\cite{Strogatz}.
While the amplitude is tightly constrained to  the limit cycle curve, the phase can vary more freely.
Hence many model studies of synchronisation have focused only on the phase dynamics~\cite{VJ2006, NEL2008,UG10,UG2011,Uchida}.
Our goal in this paper is to analyse the role played by  both the {\em amplitude} and {\em phase} dynamics on {\em phase} synchronisation mediated by  hydrodynamics. 
We first study a pair of well separated deterministic oscillators 
and find that hydrodynamic interactions strongly enhance phase locking, if the oscillations are non-isochronous, which on the limit cycle means that the frequency of oscillations depends on the amplitude.
We then consider an array of many oscillators, still well separated, in the presence of fluctuations. 
On long wavelengths  their slow %synchronisation 
dynamics  can be naturally represented in terms of a 
 broken symmetry (phase) variable, which is a  
non-equilibrium analogue of a Goldstone mode~\cite{LL-PRL}.
 Denoting by $\alpha \neq 0$ the parameter responsible for the  non-isochronity of the oscillations,
 % that on the limit cycle curve means amplitude-dependent frequency of oscillation, 
we find that when $\alpha > 0 $,  hydrodynamic interactions can  lead to  in-phase synchronisation of the array.  
These results are confirmed by numerical simulations, which show also that conversely, for $\alpha <0$,  the synchronisation  is more subtle and leads to the propagation of waves.

\paragraph{The model oscillator}
A universal model for
stable spontaneous oscillations is provided by  the normal form of a dynamical system close to a supercritical Hopf bifurcation~\cite{Strogatz}.
To be concrete, we represent the oscillator in a low Reynolds number fluid as a sphere of radius $a$ subject to a  time-varying force $f$.
The equation of motion for the sphere, with $x$ its deviation from its equilibrium position, is
\begin{equation}
 \dot{x} =  \frac{f}{\gamma}
\label{eq:x}
\end{equation} 
where $\gamma= 6 \pi \eta a$ is the Stokes drag.
The dynamics is encoded in  the evolution equation for the force $f$ : 
\begin{equation}
\dot{f} = \Psi(f,x) := -\frac{k}{\tau} x + \mu \frac{f}{\gamma} \big(1-\sigma x^2\big) + \alpha x^3 .
\label{eq:f}
\end{equation}
Here, all the parameters, except $\alpha$ 
%that can also be negative,  
are positive quantities, 
The 1st and 3rd term of eq~(\ref{eq:f}) give rise to respectively,  a linear and a non-linear {\em passive} oscillator, while the 2nd term is responsible for active, self-sustained oscillations. We emphasize that all the terms in eq~(\ref{eq:f}) would emerge naturally from coarse-graining any friction-dominated microscopic model oscillator~\cite{CLJ03,RS2005,Camalet99}. 
%%TBL
{Eqs~(\ref{eq:x}),~(\ref{eq:f}) can be conveniently non-dimensionalised
as  $\dot{x} = f$; and $\dot{f} = -x + \epsilon_\mu f (1-x^2) + \epsilon_\alpha x^3 $, choosing units where $\tau = \frac{\gamma}{k}$. They correspond to a weakly non-linear Van der Pol-Duffing oscillator~\cite{Strogatz}.
 The %non-dimensionalised 
 parameters $\epsilon_\mu:= \frac{\mu}{k}  $ and $\epsilon_\alpha:= \frac{\alpha \tau}{k \sigma}$ are small quantities. We restrict ourselves here mainly to cases where $\epsilon_\alpha/\epsilon_\mu = O(1) $.}
%By taking the time derivative of eq~(\ref{eq:x}) inserting eq~(\ref{eq:f})  and using eq~(\ref{eq:x}) to express $f$ as a function of $\dot{x}$  one obtains
%an equation in the form of a weakly non-linear Van der Pol-Duffing oscillator~\cite{Strogatz}.
%$\alpha \neq 0$  makes oscillations  non-isochronous~\cite{Pikovsky}. 

\begin{figure}[h!]
  \includegraphics[width=0.3\textwidth]{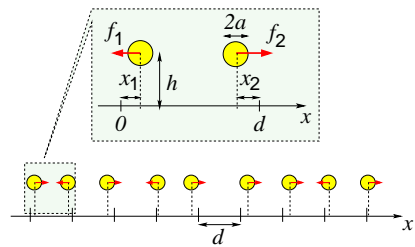}
\caption{(color online) One dimensional  lattice of oscillators. The inset
 illustrates the dynamic variables of  a pair.
\label{fig:scheme} }
\end{figure}

\paragraph{Two oscillators coupled hydrodynamically}
%We consider 
The oscillators are arranged along the x-axis. 
%We shall assume that also
 The forces $f_i$ acting on the spheres, for $i=1,2$,  are directed along the same axis
and cause  sphere 1 to oscillate around the origin and sphere 2 around position $d$.  We denote by $x_i$  the deviations from these equilibrium positions, see fig~\ref{fig:scheme}. 
 %Due to this arrangements, 
 Their equations of motion are %with hydrodynamic interactions $H(r)$   
\begin{equation}
\left\{ \begin{array}{l}
\dot{x}_1 = \frac{1}{\gamma} \left( f_1 +  H(r) f_2 \right);\\
\\
\dot{x}_2 = \frac{1}{\gamma}\left( f_2 +  H(r)  f_1\right),\\
\end{array}\right.
\label{eq:xi}
% \left( \dfrac{h}{r}\right)^\kappa
\end{equation}
where $H(r)$ is a scalar, representing the hydrodynamic interactions, and $r := d+ x_2 - x_1 $ is the separation between the sphere centres.
We shall consider the limit of large separation $r$ compared to the sphere radius $a$.
 Then, for an unbounded three-dimensional fluid,  interactions are described by  the Oseen tensor~\cite{Doi}
as $H(r)= \frac{3a}{2r}$.
For a rigid surface with a  non-slip boundary condition, placed at distance $h$ from  the oscillators,  one obtains  effective 
interactions scaling as $H(r) \sim \frac{ah^2}{r^3} $~\cite{Blake}.
For an assembly of oscillators arranged on a regular lattice,  $d$ can be thought of as the lattice spacing, see fig~\ref{fig:scheme}.
We {assume} that it is large compared to the amplitude of the oscillations, $d > x_2 -x_1$, and that the ratio {$\epsilon_d := a/d$, characterising the hydrodynamic coupling, satisfies  $ \epsilon_d \ll \epsilon_\mu, \epsilon_\alpha$}.
The time evolution of forces is 
given by $\dot{f}_i = \Psi(f_i, x_i)$, with $ \Psi(f_i, x_i)$ defined in eq~(\ref{eq:f}), and is entirely local~\cite{CLJ03}.
%\begin{equation}
%\dot{f}_i = - \frac{k}{\tau} x_i + \mu \frac{f_i}{\gamma} \big(1- \sigma x^2_i \big) +\alpha x^3_i,
%\label{eq:f-i}
%\end{equation}
The long-range hydrodynamic coupling links the coordinates $x_i$ via eq~(\ref{eq:xi}).
In the following we  denote the nonlinear parts of $\Psi(f_i, x_i)$  by 
$\mathcal{F}_i(x_i, f_i) := \frac{\mu}{\gamma} f_i \big(1-\sigma x^2_i\big) + \alpha x^3_i$.

To proceed, we take the time derivative of both sides of eq~(\ref{eq:xi}) and use, on the rhs,  the evolution equation for the forces
%~(\ref{eq:f-i}) 
and the expression of forces as functions of velocities $\dot{x}_i$ obtained by inverting eq~(\ref{eq:xi}) as an expansion in $a/r$.  Thus, to leading order, we obtain equations for oscillators with reactive couplings~\cite{Pikovsky}
(given by  $ \frac{\mathfrak{D}}{\gamma}$) as
$\ddot{x}_1 + \omega^2_0 x_1 = \frac{1 }{\gamma}\mathcal{F}_1(x_1, \gamma \dot{x}_1)    + \frac{\mathfrak{D}}{\gamma} x_2 $
and $\ddot{x}_2 + \omega^2_0 x_2 = \frac{ 1}{\gamma}\mathcal{F}_2(x_2, \gamma \dot{x}_2) + \frac{\mathfrak{D}}{\gamma}  x_1$, 
where $\mathfrak{D} := -H(d) \frac{k}{\tau}$. 
%For small oscillations, $\mathfrak{D} \approx -H(d) \frac{k}{\tau}$  is approximatively  constant.
 $\omega_0$ represents the natural  frequency of the linear oscillators,  defined by $\omega^2_0 = \frac{k}{\gamma \tau}$.
Note terms like 
$\frac{H(r)}{\gamma} \mathcal{F}_i(x_i, \gamma \dot{x}_i)  $,  { of order $ \mathcal{O}(\epsilon_d \epsilon_\mu )$ } 
and  $\frac{d H}{d r} \dot{r} f_i $, { of order $ \mathcal{O}(\epsilon^2_d )$  } have been neglected.

 We now derive the equations governing the slow dynamics of the oscillators~\cite{Pikovsky}.
%Oscillations are 
This is done naturally %described
 using a complex amplitude
 $A_k$ and its complex conjugate $A^*_k $  related to position and velocities by 
%\begin{align}
%&
$x_k = \frac{1}{2} (A_k e^{i \omega t}  + A^*_k e^{-i \omega t}  )$ %\\
%&
and  $\dot{x}_k =   \frac{i  \omega}{2} (A_k e^{i \omega t}  - A^*_k e^{-i \omega t}  )$
%\label{eq:change}
%\end{align}
for $k=1,2$. This requires of course that $\dot A^*_k = -\dot{A}_k e^{2 i \omega t}$. Here $\omega$ is  the (unknown) frequency  of the non-linear oscillators, determining the period, $T=\frac{2 \pi}{\omega}$ of the (fast) oscillations. 
 The (slow) complex amplitudes, on the other hand, hardly change on this timescale.
Writing eq~(\ref{eq:xi}) and
%~(\ref{eq:f-i})
the dynamic equations for the forces 
 in terms of  $A_k$ and $A^*_k$ and averaging over the period $T $ 
one obtains 
\begin{align}
&\dot{A}_1 = - i \Delta  A_1 + \lambda A_1 -(\beta + i \chi ) A_1 |A_1|^2  + i \delta A_2 \nonumber \\
\nonumber \\
& \dot{A}_2 = - i \Delta A_2 + \lambda A_2 -(\beta +i \chi) A_2 |A_2|^2  + i \delta A_1. \label{eq:A}
 \end{align}
The parameters are defined as
$ \Delta 
:= \frac{\omega^2- \omega^2_0}{2 \omega }  $,
 $\lambda := \frac{\mu}{2\gamma} $,
 $ \beta := \frac{\mu \sigma}{8 \gamma }  $,
  $ \chi := \frac{3}{8} \frac{\alpha}{ \gamma \omega} $, and
 $ \delta(d) :=  \frac{H(d)}{2 \gamma \omega }\frac{k}{\tau}  $.
%For Stokeslets,  $ \delta =  \frac{3 a}{4 \gamma \omega d}\frac{k}{\tau} $.
 %In presence of a no-slip boundary, as discussed before, $\delta$ would scale differently, as $ \delta \sim  \frac{ a h^2}{ \gamma \omega d^3}\frac{k}{\tau} $.

Writing the complex amplitudes $A_k$ in  polar form, $A_k = R_k e^{i \phi_k}$,
eqs~(\ref{eq:A}) become a coupled system for the amplitudes $R_k$ and the phases $\phi_k$.
Finally, this system can be reduced to a single equation for the phase difference~\cite{Pikovsky}.
This can be achieved perturbatively, when the parameter $\delta$, parametrising the hydrodynamic  interactions, is small compared to the other terms. 
If interactions are neglected, $R_k$ have  fixed points given by $R_k = \sqrt{\frac{ \lambda}{\beta} } $.
The dynamics of  small deviations from these fixed points can be studied by writing  $R_k =  \sqrt{\frac{ \lambda}{\beta} } (1 + s_k)$, for $s_k \ll 1$. 
%This leads to the pair of equations
 %  %\begin{align}
%%&
%$\frac{1}{\lambda}\dot{s}_1
%\approx  - 2 s_1   - \tilde{\delta}  \sin \psi   $
%%\\
%% &
%and
 %$\frac{1}{\lambda} \dot{s}_2
  %\approx - 2 s_2   + \tilde{\delta}  \sin \psi $
%% \end{align}
 %where $\tilde{\delta} = \frac{\delta}{\lambda}$ and $ \psi := \phi_2-\phi_1$.
%These equations  show that  
One finds that the  deviations $s_k$ %are strongly dumped, and 
relax quickly to zero.
Setting $\dot{s}_k = 0$  we obtain $s_k$ as functions of  the phase difference $\psi := \phi_2-\phi_1$.
The resulting expressions are then substituted in the  equations for the phases. From them one obtains  an Adler equation~\cite{Pikovsky}  for $\psi$,
%of the form
\begin{equation}
\dot{\psi} = \tilde{\nu} -2 \frac{ \delta  \chi  }{\beta} \sin \psi.
\label{eq:adler}
 \end{equation}
 %This describes the

Hence,  eq~(\ref{eq:adler})  illustrates  that phase locking is determined  by the hydrodynamic coupling, via $\delta$, provided the oscillator is nonisochronous, i.e.  $\alpha \neq 0$.
{Note that $\frac{ \delta  \chi  }{\beta}$ scales as  $\sim \frac{1}{\tau} \frac{\epsilon_\alpha}{\epsilon_\mu}  \epsilon_d$} and 
$\tilde{\nu}$ 
is related to the difference of the natural frequencies of the oscillators. 
%In our case, 
We choose them to be  identical, so we can set $\tilde{\nu}=0$.
While for $ \tilde{\nu} \neq 0$ varying the ratio of %the parameters 
$\tilde{\nu}$ and $\frac{ \delta  \chi  }{\beta} $ controls the saddle-node bifurcation of cycles~\cite{Strogatz}, for $\tilde{\nu}=0$  eq~(\ref{eq:adler}) has a stable fixed point given by one of the zeros of  $\sin(\psi)$ for $\psi \in [0, 2 \pi]$.
The position of the stable point
 is determined by the sign of  $ -\frac{ \delta  \chi  }{\beta}$, which in turn is determined solely by the sign of the non-isochronism parameter $\alpha$:
  when $\alpha < 0$, then the equation has a stable fixed point at $\psi =\pi$, i.e. the oscillators lock in anti-phase; vice-versa, if $ \alpha > 0 $ then the equation has a stable fixed point at $\psi =0$ and the oscillators lock in-phase.
A numerical solution, using the  Euler method, of eq~(\ref{eq:xi}) confirms this.

%It is possible to make an analogy with spin models of magnets:
 % the case $\alpha > 0$ corresponds to a ferromagnetic coupling, favouring alignment of neighbour spins, 
 %whereas $\alpha < 0 $ can be viewed  as an anti-ferromagnetic coupling, leading to anti-alignment.

It is also interesting to note that the two flagella of the microscopic 
algae \emph{C. Reinhardtii}
are found to alternate between periods of  synchronised  (with small phase difference)  and non-synchronized beating~\cite{Polin, Gold}. This is well described by a stochastic Adler equation, of the same form as  eq~(\ref{eq:adler}) but with an additional fluctuating term~\cite{Gold}.
 % The noise  represents the effect of thermal (and presumably active) fluctuations in the phases of the two oscillators. %, which at the micro-scale are important.
 The estimates of the parameters presented in~\cite{Gold}, for the flagellar synchronisation,  indicate positive values for   $\alpha$ and $\tilde{\nu}$  of our model.
 
 When  $\alpha=0$,
we need to include  higher order corrections in deriving eq~(\ref{eq:adler}). Upon doing this 
we find to leading order 
%{\color{blue} (ML) the non-dimensional equation }
$\dot{\psi} \approx -3 \epsilon_d   \epsilon_\mu [ 1  + \frac{3}{4} \frac{\epsilon_d}{\epsilon^2_\mu} \cos\psi ]  \sin  \psi.$ 
When  $\epsilon_d < \frac{4}{3} \epsilon^2_\mu $  % the fixed points are determined  by $\sin \psi =0$ alone and
 the synchronisation is in-phase.
{Otherwise, both in- and anti-phase states are possible and synchronization depends on details such as initial conditions (confirmed numerically).}
These higher order terms also indicate that  the transition from in-phase to anti-phase in general occurs at some $\alpha_c  \ne 0$.
Unsurprisingly when $\alpha =0$, synchronisation occurs more slowly (a higher order effect).
% but is shifted to a negative value.
%However, for large enough values of $\alpha$ there is symmetry between these states.
%, set by the inverse of $\epsilon_d   \epsilon_\mu$, 
%compared to that of eq~(\ref{eq:adler}), as expected from high order terms. % where $\epsilon_\alpha \sim \epsilon_\mu$

\paragraph{Many oscillators coupled hydrodynamically}
As we have discussed above, the amplitudes of the oscillators are tightly constrained to the limit cycle
%. They are not relevant variables 
and the long time behaviour can be reduced to an effective (amplitude dependent) dynamics of the phases.  
For a large number $N$ of oscillators, in the dilute regime, this  is done by introducing the one-particle probability
%In the dilute limit, we introducing the  one-particle probability 
$c(\varphi, y, t ) = \langle \frac{1}{N} \sum^{N}_{k=1} \delta(\varphi -\phi_k(t) ) \, \delta( y-y_k(t)) \rangle$ of having an oscillator with slow phase $\varphi$, at site $y$
at time $t$, where the brackets  $\langle \rangle$ indicate the average over noise.
The probability satisfies a Smoluchowski equation 
\begin{equation}
\partial_t c = D \partial^2_{\varphi  \varphi } c - \partial_\varphi( [ \omega_1 +\Omega] c  ).
\label{eq:smol}
 \end{equation}
   $D $ is the diffusion coefficient resulting from both thermal and active fluctuations, $\omega_1$  the deterministic contribution of an isolated oscillator
with 
 $\omega_1 = -\Delta -\frac{\chi \lambda}{\beta}$ %-  \frac{ \chi \lambda }{\beta}$
    and $\Omega$  the deterministic effect of the  hydrodynamic interactions, % defined as
\begin{equation}
\Omega(y, \varphi, t) := \int d y_2 d \varphi_2 c(  \varphi_2, y_2,  t) \dot{\phi}^{int}( y_2-y, \varphi, \varphi_2).
\label{eq:Omega}
\end{equation}
 { $\dot{\phi}^{int} =  \frac{\chi \delta' }{ \beta }\sin( \varphi_2 -\varphi)  +\delta'  \cos( \varphi_2-\varphi)  $}
 is obtained from the dynamics of two oscillators, 
(see  eq~(\ref{eq:adler})). 
It describes the effect of the interactions on the phase of one oscillator due to the presence of the another. Here, $\delta' := \delta(|y_2-y|) $.

The 1-particle probability can be expressed as an expansion in its moments: 
%This in turn can be given as a function its  moments. Truncating after the second one, we find 
\begin{equation}
c(\varphi, y, t) =  \frac{1}{2 \pi }\left[ \rho(y, t) +  \left(e^{-i\varphi} \Phi(y, t)  + \mbox{c.c.} \right)+ \ldots\right]
\label{eq:trunc}
\end{equation}
To study synchronization we only need the first two :
%The first two moments of the 1-particle probability are: 
\begin{align}
&\rho(y, t) := \int^{2 \pi}_0 d\varphi c(\varphi, y, t) \;  ; \quad \mbox{(density)}\nonumber \\
& \Phi(y, t) := \int^{2 \pi}_0 d\varphi e^{i \varphi} c( \varphi, y, t) \; ; \quad \mbox{(1st harmonic)} \; .
\label{eq:moments}
\end{align}
%as the density and the first harmonic of the probability. 

%which we truncate after the 2nd.

The emergence (or not) of a globally synchronized state is obtained from  the homogeneous probability $c^0( \varphi,t) $, with associated moments $\rho^0(t)$, $\Phi^0(t)$ representing spatially homogeneous dynamical states. 
The corresponding expression  for  $\Omega^0$ is obtained by  evaluating the space integral in eq~(\ref{eq:Omega}) with $c \equiv c^0$. For hydrodynamic interactions scaling as $H(r) \sim \frac{a}{r}$ the leading term from the  integral depends both on the  lattice spacing $d$, and the total length $L$ of the array.
%\begin{align}
%&
Hence,
{$\Omega^0(t) = \frac{3a k}{4\omega \tau \gamma} \ln(L/d)  [ - i  \frac{\chi}{\beta}    + 1]  e^{-i \varphi_1}  \Phi^0(t) + \mbox{c.c.} $ }
For interactions scaling as $H(r) \sim \frac{a h^2}{r^3}$, the leading term in the integral depends only on  the lattice spacing $d$. Consequently, the term $a \ln{(L/d)}$  is replaced by one $ \sim \frac{a h^2}{d^2} $.
%\end{align}
Dynamic equations for the homogeneous moments are derived by taking the time derivative of both sides of eq~(\ref{eq:moments}), inserting eq~(\ref{eq:smol}) and using  eq~(\ref{eq:trunc}) to close the system.
 Since $\rho$ is a conserved variable,
 $ \partial_t \rho^0 = 0$,
while $\Phi^0$
satisfies 
\begin{align}
\partial_t \Phi^0 = \Gamma   \Phi^0.
\label{eq:phi0}
\end{align}

It is worth noting that in the  absence of noise  $c^0(\varphi, t ) = \frac{1}{N} \sum^N_{k=1} \delta(\varphi -\varphi_k(t))$ and
$\Phi^0(t)$ reduces to the order parameter introduced by Kuramoto, $\Phi^0(t) = \frac{1}{N}  \sum^N_{k=1} e^{i \varphi_k(t) }  $, representing the (mean field)  average over  a  population of oscillators~\cite{Pikovsky, Ritort,Uchida}.
%%{\color{red} ??? Hence, eq~(\ref{eq:phi0}) describes its dynamics, during the synchronisation transition.}
%However, here unlike the Kuramoto model, the interaction strength depends on distance between the oscillators~\cite{Uchida}.

It is useful to express $\Phi^0(t) = P^0(t) e^{i Q^0(t)}$ in polar form (reflecting the $U(1)$ symmetry).
%The underlying $U(1)$ symmetry means  we can express  $\Phi^0(t) = P^0(t) e^{i Q^0(t)}$ in polar form.
%By multiplying the equations for $\Phi^0$ and  for $(\Phi^0)^*$ respectively by $e^{i Q^0}$ and  $e^{-i Q^0}$,   adding and subtracting both sides of the resulting equations, eq~(\ref{eq:phi0})  decouples in 
%\begin{align}
%&
We obtain equations for its amplitude and phase as
$ \partial_t P^0  =  Re[\Gamma] P^0 $
and
% \nonumber \\
%&
 $\partial_t Q^0  =  Im[\Gamma]$. 
%\end{align}
$Re(\Gamma) = -( D -\frac{\chi}{\beta}  \frac{3a k}{4\omega \tau \gamma} \ln(L/d) \rho^0 ) $ is the real part of $\Gamma$. Here, the first term is due to noise, whereas the second term encodes the effect of two body interactions.
The imaginary part  is
{$Im(\Gamma) =[ \omega_1 +  \frac{3a k}{4\omega \tau \gamma} \ln(L/d)\rho^0   ] $. }

\begin{figure}[h!]
  \includegraphics[width=0.48\textwidth]{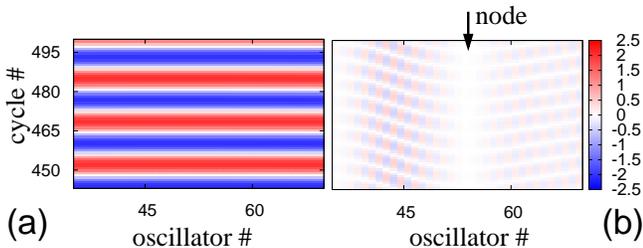}
\caption{ (color online)  Space-time plots of the positions for $N=100$ deterministic oscillators, ($D=0$).  
(a), (b) describe  respectively the case for $\alpha >0$ and $\alpha <0$,
after  long time. 
The initial conditions of the oscillators are the same for both values of $\alpha$:  identical amplitudes, close to the maximum value, and random, Gaussian distributed, phases.  
 The parameters of the model are $\gamma =10^{-3}  Pa \left. \right. s \left. \right. \mu m  $;   $\frac{k}{\tau} = 1 \frac{pN}{\mu m \left. \right. s} $;  $\mu = 0.05  \frac{pN}{\mu m} $;  $\sigma = 1 (\mu m)^{-2}$;  $ |\alpha | =  0.05 \frac{pN}{(\mu m)^3 s}$ and  $a/d \approx 0.005$.
\label{fig:spacetime} }
\end{figure}

As in the Kuramoto model~\cite{Pikovsky, Ritort},  order (synchronisation)  is determined by a non-zero, constant value of  $P^0$. 
Here,  the dynamic equation for $P^0$ shows that the onset of order is controlled by  the sign of $Re[\Gamma] $.
If $Re[\Gamma] <0$, 
 %$P^0$ that  relaxes to zero exponentially, so 
  order is suppressed.
On the contrary, when  $Re[\Gamma] > 0$,  order is enhanced. %As it stands, the equation for $P^0$ describes exponential growth in time. 
A  stabilising term of the type $\sim \Phi^0 |\Phi^0|^2$ in eq~(\ref{eq:phi0}) is needed for $P^0$ to stop unbounded growth and attain a  finite value at long times. Such a term could be generated for instance by taking into account three-body interactions.
Finally, the condition $Re[\Gamma]  =0 $ 
 defines a transition line in the space of parameters~\cite{LL-PRL}.
Crucially, from these considerations, homogeneous synchronization is possible only when $\alpha >0 $:
(i) in presence of noise ($D \neq 0$) and by keeping all the  parameters fixed, %one can easily see that 
synchronisation occurs only above a particular value of density; (ii)  neglecting noise ($D=0$), instead, synchronisation occurs for any (finite) value of the density.
On the contrary, when $\alpha <0$ both terms in $Re(\Gamma)$ are negative and homogeneous order is prohibited.
This behaviour suggests a spin  analogy,  where $\alpha >0$ (ferromagnet) promotes alignment of neighbouring oscillator phase (spins) while $\alpha<0$ (antiferromagnet) promotes anti-alignment.

We compared these results with numerical simulation for a large but finite number of deterministic oscillators ($D=0$). In fig~\ref{fig:spacetime} we show  typical space-time plots for the positions of $N=100$ oscillators and compare the effects of different signs of $\alpha$.
For $\alpha > 0$, see fig~\ref{fig:spacetime}(a),  the system displays  spatially homogeneous order, i.e. in-phase synchronised state. Interestingly, when $\alpha < 0$,  although homogeneous order is lacking, fig~\ref{fig:spacetime}(b) still shows  a coherent motion of the oscillators, with propagating waves. 
As suggested by the antiferromagnetic analogy,  the oscillators self-organise into a dynamical state which is close to the anti-phase synchronised state, but deviates from it at long wavelengths.

In conclusion, we have presented a simple, one-dimensional model 
{(that can be generalised to higher dimensions~\cite{LL12})}
 to investigate analytically the role of hydrodynamic interactions on the synchronisation dynamics of oscillators at low Reynolds number.
We studied  the case of two oscillators and found that synchronisation, either in- or anti-phase, was determined to leading order by both hydrodynamic interactions and non-isochronism of the oscillations ($\alpha \neq 0$).
We then derived a coarse grained description for an infinite array of oscillators and found that spatially homogeneous order, corresponding to the in-phase synchronisation of the array, can occur only  for $\alpha > 0$. 
%Such condition is indicated by matching the estimates of %performed on 
%recent in-vivo studies~\cite{Gold} with our model. 
%Very recent studies~\cite{Hamel} revealed that coordinated in-phase motion of cilia may be adopted by  Paramecium as a strategy of swimming.
%However, 
Systems of cilia are known to display metachronal waves~\cite{Hamel}. 
Our analysis suggests that  these could be obtained in two different ways: either as slow hydrodynamic (phase) modes, like spin waves, when $\alpha >0$; or
alternatively,  for $\alpha <  0$, as a spatially inhomogeneous, approximately anti-phase synchronised state, as indicated by the numerics.
A more extensive investigation of these issues is left for the future.  
 
We acknowledge the support of the EPSRC, Grant EP/G026440/1 (ML \& TBL);  the NSF, Grant PHY05-51164 (TBL);  and the University of Bristol (ML).
 \bibliography{synch}

\begin{thebibliography}{24}
\expandafter\ifx\csname natexlab\endcsname\relax\def\natexlab#1{#1}\fi
\expandafter\ifx\csname bibnamefont\endcsname\relax
  \def\bibnamefont#1{#1}\fi
\expandafter\ifx\csname bibfnamefont\endcsname\relax
  \def\bibfnamefont#1{#1}\fi
\expandafter\ifx\csname citenamefont\endcsname\relax
  \def\citenamefont#1{#1}\fi
\expandafter\ifx\csname url\endcsname\relax
  \def\url#1{\texttt{#1}}\fi
\expandafter\ifx\csname urlprefix\endcsname\relax\def\urlprefix{URL }\fi
\providecommand{\bibinfo}[2]{#2}
\providecommand{\eprint}[2][]{\url{#2}}

\bibitem[{\citenamefont{Pikovsky et~al.}(2002)\citenamefont{Pikovsky,
  Rosenblum, and Kurths}}]{Pikovsky}
\bibinfo{author}{\bibfnamefont{A.}~\bibnamefont{Pikovsky}},
  \bibinfo{author}{\bibfnamefont{M.}~\bibnamefont{Rosenblum}},
  \bibnamefont{and} \bibinfo{author}{\bibfnamefont{J.}~\bibnamefont{Kurths}},
  \emph{\bibinfo{title}{Synchronization: A Universal Concept in Nonlinear
  Science}} (\bibinfo{publisher}{{Cambridge University Press}},
  \bibinfo{year}{2002}).

\bibitem[{\citenamefont{Bray}(2000)}]{Bray}
\bibinfo{author}{\bibfnamefont{D.}~\bibnamefont{Bray}},
  \emph{\bibinfo{title}{Cell Movements: From Molecules to Motility}}
  (\bibinfo{publisher}{Garland Science, New York}, \bibinfo{year}{2000}).

\bibitem[{\citenamefont{Nonaka et~al.}(1998)}]{Nonaka}
\bibinfo{author}{\bibfnamefont{S.}~\bibnamefont{Nonaka}} \bibnamefont{et~al.},
  \bibinfo{journal}{Cell} \textbf{\bibinfo{volume}{95}}, \bibinfo{pages}{829}
  (\bibinfo{year}{1998}).

\bibitem[{\citenamefont{Lagomarsino et~al.}(2003)\citenamefont{Lagomarsino,
  Jona, and Bassetti}}]{CLJ03}
\bibinfo{author}{\bibfnamefont{M.~C.} \bibnamefont{Lagomarsino}},
  \bibinfo{author}{\bibfnamefont{P.}~\bibnamefont{Jona}}, \bibnamefont{and}
  \bibinfo{author}{\bibfnamefont{B.}~\bibnamefont{Bassetti}},
  \bibinfo{journal}{Phys. Rev. E} \textbf{\bibinfo{volume}{68}},
  \bibinfo{pages}{021908} (\bibinfo{year}{2003}).

\bibitem[{\citenamefont{Vilfan and Julicher}(2006)}]{VJ2006}
\bibinfo{author}{\bibfnamefont{A.}~\bibnamefont{Vilfan}} \bibnamefont{and}
  \bibinfo{author}{\bibfnamefont{F.}~\bibnamefont{Julicher}},
  \bibinfo{journal}{Phys. Rev. Lett.} \textbf{\bibinfo{volume}{96}},
  \bibinfo{pages}{058102} (\bibinfo{year}{2006}).

\bibitem[{\citenamefont{Niedermayer et~al.}(2008)\citenamefont{Niedermayer,
  Eckhardt, and Lenz}}]{NEL2008}
\bibinfo{author}{\bibfnamefont{T.}~\bibnamefont{Niedermayer}},
  \bibinfo{author}{\bibfnamefont{B.}~\bibnamefont{Eckhardt}}, \bibnamefont{and}
  \bibinfo{author}{\bibfnamefont{P.}~\bibnamefont{Lenz}},
  \bibinfo{journal}{Chaos} \textbf{\bibinfo{volume}{18}},
  \bibinfo{pages}{037128} (\bibinfo{year}{2008}).

\bibitem[{\citenamefont{Uchida and Golestanian}(2010)}]{UG10}
\bibinfo{author}{\bibfnamefont{N.}~\bibnamefont{Uchida}} \bibnamefont{and}
  \bibinfo{author}{\bibfnamefont{R.}~\bibnamefont{Golestanian}},
  \bibinfo{journal}{Phys. Rev. Lett.} \textbf{\bibinfo{volume}{104}},
  \bibinfo{pages}{178103} (\bibinfo{year}{2010}).

\bibitem[{\citenamefont{Guirao and Joanny}(2007)}]{GJ07}
\bibinfo{author}{\bibfnamefont{B.}~\bibnamefont{Guirao}} \bibnamefont{and}
  \bibinfo{author}{\bibfnamefont{J.~F.} \bibnamefont{Joanny}},
  \bibinfo{journal}{Biophys. J.} \textbf{\bibinfo{volume}{92}},
  \bibinfo{pages}{1900} (\bibinfo{year}{2007}).

\bibitem[{\citenamefont{Kim and Powers}(2004)}]{KP04}
\bibinfo{author}{\bibfnamefont{M.}~\bibnamefont{Kim}} \bibnamefont{and}
  \bibinfo{author}{\bibfnamefont{T.~R.} \bibnamefont{Powers}},
  \bibinfo{journal}{Phys. Rev. E} \textbf{\bibinfo{volume}{69}},
  \bibinfo{pages}{061910} (\bibinfo{year}{2004}).

\bibitem[{\citenamefont{Reichert and Stark}(2005)}]{RS2005}
\bibinfo{author}{\bibfnamefont{M.}~\bibnamefont{Reichert}} \bibnamefont{and}
  \bibinfo{author}{\bibfnamefont{H.}~\bibnamefont{Stark}},
  \bibinfo{journal}{Eur. Phys. J. E} \textbf{\bibinfo{volume}{17}},
  \bibinfo{pages}{493} (\bibinfo{year}{2005}).

\bibitem[{\citenamefont{Uchida and Golestanian}(2011)}]{UG2011}
\bibinfo{author}{\bibfnamefont{N.}~\bibnamefont{Uchida}} \bibnamefont{and}
  \bibinfo{author}{\bibfnamefont{R.}~\bibnamefont{Golestanian}},
  \bibinfo{journal}{Phys. Rev. Lett.} \textbf{\bibinfo{volume}{106}},
  \bibinfo{pages}{058104} (\bibinfo{year}{2011}).

\bibitem[{\citenamefont{Uchida}(2011)}]{Uchida}
\bibinfo{author}{\bibfnamefont{N.}~\bibnamefont{Uchida}},
  \bibinfo{journal}{Phys. Rev. Lett.} \textbf{\bibinfo{volume}{106}},
  \bibinfo{pages}{064101} (\bibinfo{year}{2011}).

\bibitem[{\citenamefont{Polin et~al.}(2009)}]{Polin}
\bibinfo{author}{\bibfnamefont{M.}~\bibnamefont{Polin}} \bibnamefont{et~al.},
  \bibinfo{journal}{Science} \textbf{\bibinfo{volume}{325}},
  \bibinfo{pages}{487} (\bibinfo{year}{2009}).

\bibitem[{\citenamefont{Goldstein et~al.}(2009)\citenamefont{Goldstein, Polin,
  and Tuval}}]{Gold}
\bibinfo{author}{\bibfnamefont{R.~E.} \bibnamefont{Goldstein}},
  \bibinfo{author}{\bibfnamefont{M.}~\bibnamefont{Polin}}, \bibnamefont{and}
  \bibinfo{author}{\bibfnamefont{I.}~\bibnamefont{Tuval}},
  \bibinfo{journal}{Phys. Rev. Lett.} \textbf{\bibinfo{volume}{103}},
  \bibinfo{pages}{168103} (\bibinfo{year}{2009}).

\bibitem[{\citenamefont{Kotar et~al.}(2010)}]{kotar}
\bibinfo{author}{\bibfnamefont{J.}~\bibnamefont{Kotar}} \bibnamefont{et~al.},
  \bibinfo{journal}{PNAS} \textbf{\bibinfo{volume}{107}}, \bibinfo{pages}{7669}
  (\bibinfo{year}{2010}).

\bibitem[{\citenamefont{Qian et~al.}(2009)}]{Qian}
\bibinfo{author}{\bibfnamefont{B.}~\bibnamefont{Qian}} \bibnamefont{et~al.},
  \bibinfo{journal}{Phys. Rev. E} \textbf{\bibinfo{volume}{80}},
  \bibinfo{pages}{061919} (\bibinfo{year}{2009}).

\bibitem[{\citenamefont{Camalet et~al.}(1999)\citenamefont{Camalet, Julicher,
  and Prost}}]{Camalet99}
\bibinfo{author}{\bibfnamefont{S.}~\bibnamefont{Camalet}},
  \bibinfo{author}{\bibfnamefont{F.}~\bibnamefont{Julicher}}, \bibnamefont{and}
  \bibinfo{author}{\bibfnamefont{J.}~\bibnamefont{Prost}},
  \bibinfo{journal}{Phys. Rev. Lett.} \textbf{\bibinfo{volume}{82}},
  \bibinfo{pages}{1590} (\bibinfo{year}{1999}).

\bibitem[{\citenamefont{Strogatz}(2001)}]{Strogatz}
\bibinfo{author}{\bibfnamefont{S.}~\bibnamefont{Strogatz}},
  \emph{\bibinfo{title}{Nonlinear Dynamics And Chaos}}
  (\bibinfo{publisher}{Westview press}, \bibinfo{year}{2001}).

\bibitem[{\citenamefont{Leoni and Liverpool}(2010)}]{LL-PRL}
\bibinfo{author}{\bibfnamefont{M.}~\bibnamefont{Leoni}} \bibnamefont{and}
  \bibinfo{author}{\bibfnamefont{T.~B.} \bibnamefont{Liverpool}},
  \bibinfo{journal}{Phys Rev Lett} \textbf{\bibinfo{volume}{105}},
  \bibinfo{pages}{238102} (\bibinfo{year}{2010}).

\bibitem[{\citenamefont{Doi and Edwards}(1986)}]{Doi}
\bibinfo{author}{\bibfnamefont{M.}~\bibnamefont{Doi}} \bibnamefont{and}
  \bibinfo{author}{\bibfnamefont{S.}~\bibnamefont{Edwards}},
  \emph{\bibinfo{title}{The Theory of Polymer Dynamics}}
  (\bibinfo{publisher}{Oxford University Press}, \bibinfo{year}{1986}).

\bibitem[{\citenamefont{Blake}(1971)}]{Blake}
\bibinfo{author}{\bibfnamefont{J.~R.} \bibnamefont{Blake}},
  \bibinfo{journal}{Proc. Cambridge Philos. Soc.}
  \textbf{\bibinfo{volume}{70}}, \bibinfo{pages}{303} (\bibinfo{year}{1971}).

\bibitem[{\citenamefont{Acebr\'on et~al.}(2005)}]{Ritort}
\bibinfo{author}{\bibfnamefont{J.~A.} \bibnamefont{Acebr\'on}}
  \bibnamefont{et~al.}, \bibinfo{journal}{Rev. Mod. Phys.}
  \textbf{\bibinfo{volume}{77}}, \bibinfo{pages}{137} (\bibinfo{year}{2005}).

\bibitem[{\citenamefont{Leoni and Liverpool}(2012)}]{LL12}
\bibinfo{author}{\bibfnamefont{M.}~\bibnamefont{Leoni}} \bibnamefont{and}
  \bibinfo{author}{\bibfnamefont{T.~B.} \bibnamefont{Liverpool}},
  \bibinfo{journal}{unpublished}  (\bibinfo{year}{2012}).

\bibitem[{\citenamefont{Hamel et~al.}(2011)}]{Hamel}
\bibinfo{author}{\bibfnamefont{A.}~\bibnamefont{Hamel}} \bibnamefont{et~al.},
  \bibinfo{journal}{Proc Natl Acad Sci USA} \textbf{\bibinfo{volume}{108}},
  \bibinfo{pages}{7290} (\bibinfo{year}{2011}).

\end{thebibliography}
 
\end{document}